\documentclass[12pt,a4paper]{article}
\usepackage{t1enc}
\usepackage[latin1]{inputenc}
\usepackage[english]{babel}
\normalfont
\usepackage{yfonts}
\usepackage{bbm}
\usepackage{bm}
\usepackage{wasysym}
\usepackage{amssymb}
\usepackage{graphics}
\usepackage{cite}
\usepackage{fancyhdr}

\begin{document}

\title{D-branes in Lie groups of rank $> 1$}
\author{S.Monnier \\ \\ \small Université de Genève, Section de Mathématiques, \\ \small 2-4 Rue du Lièvre, 1211 Genève 24, Switzerland \\ \small samuel.monnier@math.unige.ch}
\date{}
\maketitle
\begin{abstract}
We consider a low-energy effective action for the gauge field on Wess-Zumino-Witten D-branes in a compact simple Lie group, in the limit $k \rightarrow \infty$. We prove that the effective action is bounded from below, and study stability of various D-brane configurations, including some class of non-maximally symmetric ones. We show that for Lie groups of rank higher than one, the D-brane ground state breaks the Kac-Moody symmetry of the boundary theory. We then give arguments hinting that the "fuzzy sphere" D2-brane which is known to be the stable brane configuration in the case of SU(2), may also correspond to the ground state in other compact simple Lie groups.
\end{abstract}

D-brane physics in background fields, giving rise to non-commutative geometry on the brane worldsheet, has been given much attention for a few years. In particular, the compactification of $d$ space-time dimensions on a compact Lie group $G$ of dimension $d$ implies a non-vanishing B-field. The degrees of freedom corresponding to these compactified dimensions are incorporated into string theory by a level $k$ Wess-Zumino-Witten model.

D-branes are characterized by gluing conditions at the boundary of the conformal field theory describing the string vacuum. In particular, maximally symmetric D-branes are defined by the gluing condition $J = \bar{J}$ on WZW boundary currents (for a discussion of more general boundary conditions, see \cite{fil6, fil7, twistD}).  Classically, it was shown \cite{fil9, fil8} that maximally symmetric D-branes coincide with conjugacy classes of the Lie group, and after quantization, D-branes become non-commutative spaces \cite{bdbf, open_string_and..., fil4, fil5}. These D-branes are stabilized at quantized radii, as a consequence of the quantization of the U(1) gauge field strength flux on the brane \cite{fil15}.

The only case of immediate physical interest, and the only case which has been studied extensively, is $G = \mbox{SU}(2)$. $S^3 \cong$ SU(2) appears in the exact supersymmetric string background $S^3 \times AdS_3$ \cite{ads3, ads3_2}, and the CFT formulation of the Neveu-Schwarz 5-brane contains a SU(2) WZW model (see e.g. \cite{ns5}). For $G =$ SU(2), the classical maximally symmetric D-branes are 2-spheres, except for two degenerate points at $\pm e$ ($e$ being the unit element of $\mbox{SU}(2)$). The corresponding non-commutative spaces are fuzzy spheres, which can be seen as truncations of the usual algebra of functions on the 2-sphere \cite{fuz_sphe}. Our aim is to study what happens with higher rank compact simple groups.

In section \ref{eff_act}, we review some material from \cite{bdbf} and \cite{Diel_branes}, give the expression for the effective action found in these papers and point out three obvious types of solutions of the equations of motion. In section \ref{ex_min}, we prove that the action is bounded from below. Then, in section \ref{sa_rep}, we give a formula computing the value of the action for maximally symmetric solutions of the equations of motion, and study their stability. Next, in section \ref{sub_rep}, we compute again the action, this time for certain symmetry breaking solutions. We prove in section \ref{stab_srep} that they are local minima. In section \ref{exam}, we specialize the discussion to $G =$ SU(3) and show with a few examples that, contrary to the case $G =$ SU(2), the global minimum is \emph{not} maximally symmetric. We finally give an argument showing that this should hold true in SU($N$), and we conjecture that the global minimum may be the famous fuzzy sphere D2-brane.

\section{ The effective action}

\label{eff_act}

In \cite{bdbf}, the authors considered branes on a Lie group and computed the low-energy effective action in leading order in the inverse of string tension for the massless open string modes. For (bosonic) string theories with $n$ Chan-Patton degrees of freedom, generating a $\mbox{U}(n)$ gauge symmetry, these string modes are described by a $n\times n$ matrix valued gauge vector field $A$. It was shown in \cite{bdbf} that all the physics can be seen in constant solutions, so we will study only these. Constant vector fields on a Lie group $G$ with Lie algebra $\textfrak{g}$ are given by their components in a basis $\{e_a\}$ of $\textfrak{g}$ and we will choose this basis to be orthonormal with respect to the Killing form. It is useful to normalize it so that the squared norm of the long roots of the Lie algebra is equal to 2 : 
$$ 
(e_a,e_b)= \frac{1}{2h^\vee} \mbox{Tr}(\mbox{ad}_{e_a} \mbox{ad}_{e_b}) = \delta_{ab} \;\; ,
$$
where $h^\vee$ is the dual Coxeter number of the Lie algebra $\textfrak{g}$.
The effective action is then given by \nolinebreak:
\begin{equation}
  \label{def_S}
  S(A) = \frac{1}{4(h^\vee)^2} \left (-\frac{1}{4} \sum_{a,b} \mbox{Tr}([A_a,A_b]^2) + \frac{1}{3} \sum_{a,b,c} f_{abc} \mbox{Tr}(A_a[A_b,A_c]) \right ) \;\;,
\end{equation}
where $A_a=A(e_a)$ are the components of the $n\times n$ matrix valued gauge field associated with the generators $\{e_a\}$. The factor $\frac{1}{4(h^\vee)^2}$ does not appear in \cite{bdbf}, and is due to our normalization of the Killing form. This action can be used for probing stability of brane configurations described by the gauge field $A$. Note that it is valid when the Regge slope $\alpha'$ goes to $0$, while the product $\alpha' k$ approaches $\infty$.

The very same action appeared in \cite{Diel_branes}, with a different interpretation. In this paper, the author considered $n$ D0-branes in a constant R-R 4-form field strength $F^{(4)} = dC^{(3)}$ with non-trivial components $F^{tijk} = -2f\epsilon_{ijk}$, $i,j,k \in {1,2,3}$ being proportional to the structure constants of su(2). The low-energy effective action found for the transverse coordinates $\Phi^i$ of the $n$ D0-branes is completely analogous to (\ref{def_S}) if we identify $\epsilon_{ijk}$ with $f_{abc}$ and $A$ with $\gamma \Phi$, with some suitable constant $\gamma$. Higher order corrections in $\alpha'$ to the effective action were computed in \cite{action_o_sup}, but we will stick to the simple action (\ref{def_S}).

This effective action is composed of a quartic Yang-Mills term and a cubic Chern-Simon term. The quartic term is always positive and appears alone in the flat space case \cite{fil1}, for which any set of commuting matrices obviously reaches the global minimum. The Chern-Simon term appearing in curved space (non-commutative Lie groups) makes the global minimum much less obvious. As already noted in \cite{bdbf} and \cite{Diel_branes}, the equations of motion are given by \nolinebreak:
\begin{equation}
  \label{first_var}
  \left ( \frac{\delta S}{\delta A} \right ) _b = \frac{1}{4(h^\vee)^2} \sum_a \left [ A_a, \left ([A_a, A_b] - \sum_c f_{abc}A_c \right ) \right ] = 0 
\end{equation}
and there are two obvious types of solutions for $A$ :
\begin{itemize}
	\item (Type I) Sets of commuting matrices : These matrices can be diagonalized, and describe $n$
D0-branes, which positions in space are given by the vector valued eigenvalues of the matricial vector $(A_a)$. If the eigenvalues are all different, the gauge symmetry $U(n)$ is broken to $U(1)^n$.
	\item (Type II) Representations of $\textfrak{g}$ ($[A_a, A_b]= \sum_c f_{abc} A_c$) : It was shown \cite{bdbf, Diel_branes} that for $G = \mbox{SU}(2)$, these solutions describe unstable stacks of D0-branes, which condense into one or more D2-branes. Their geometry is described by the "fuzzy sphere" non-commutative space \cite{fuz_sphe}, and their sizes are determined by the spin of the irreducible subrepresentations of the representation $A_a$. In particular, when the representation is irreducible, the final state consists of a single D2-brane and the gauge group is then completely broken to U(1). For groups of higher rank, fuzzy sphere analogues are fuzzy conjugacy classes.
\end{itemize}
On Lie groups of rank higher than one, there is a third type of obvious solutions (Type III), namely representations of subalgebras. Suppose for instance that the generators $e_1, e_2, e_3$ span a su(2) subalgebra of $\textfrak{g}$. We can choose the corresponding matrix components of A to form a representation of the su(2) subalgebra and set the other ones to zero, ie :
$$
[A_i, A_j] = f_{ijk}A_k \;\;\;i,j,k \in \{1,2,3\}
$$ 
and 
$$
A_s = 0\;\;\;s \neq 1, 2, 3 \;\;\;.
$$
Then (\ref{first_var}) is automatically satisfied. These solutions describe fuzzy sphere D2-branes extending in the directions 1, 2 and 3, and centered at the origin. As above, there is one D2-brane per irreducible su(2) subrepresentation. Such solutions break the Kac-Moody symmetry of the boundary currents $J$ and $\bar{J}$, so the corresponding D-branes are not maximally symmetric. 

Note that we can add to any solution $\{A_a\}$ of (\ref{first_var}) a vector $\{T_a\}$ of matrices commuting with each other and with all $A_a$'s, and still get a solution of the equations of motion. As the $T_a$'s commute with each other, they can be diagonalized. And in the case where $A$ forms a representation of either a subalgebra or the full algebra, they are proportional to identity on each irreducible submodule by Schur's Lemma. As one D-brane is associated with each irreducible subrepresentation, this freedom can be interpreted as independent shifts of the D-branes on the Lie group. 

It will also be useful to note two further symmetry properties of the action. It is invariant under the action of U($n$) by conjugation on the target matrices : 
$$
A(e_a) \mapsto U A(e_a)U^{-1} \;\;, U \in \mbox{U}(n)\;,
$$
and also under the action of G on the Lie algebra by conjugation :
$$
A(e_a) \mapsto A(ge_ag^{-1}) \;,\; g \in G \;.
$$

\vspace{.5cm}

In the case $G = \mbox{SU}(2)$ it was proved in \cite{princ_var_rep_th} that for any $n$, the global minimum of the action is realized if and only if $A$ is an irreducible representation of $\mbox{su}(2)$. It was also conjectured that for all compact simple Lie groups, the action is minimal on irreducible representations of the group Lie algebra, i.e. that the brane ground state is a fuzzy conjugacy class of $G$. 

It is difficult to get informations about the global minimum of (\ref{def_S}) directly, but the stability of all type II solutions and of some type III ones can be tested. Combined with explicit computations of $S$ on them, this will allow us to draw conclusions about the global minimum, and in particular to show that it is generally not a representation of $\textfrak{g}$. Physically the corresponding stable brane configuration is not a fuzzy conjugacy class of $G$, and therefore is not maximally symmetric.

We emphasize that this effective action, as well as the fuzzy conjugacy class picture, is valid only for $k \rightarrow \infty$. So the D-branes described here are small compared to the radius of curvature of the group manifold, or equivalently, they are localized close to the identity element of the group.

\section{Boundedness of the action}

\label{ex_min}

We would like to prove first that $S(A)$ is bounded from below. This amounts to studying its asymptotic behavior. In this section we temporarily forget the global factor $\frac{1}{4(h^\vee)^2}$ and we decompose $S(A)$ into its homogeneous part of degree four $S_4(A) = -\frac{1}{4} \sum_{a,b} \mbox{Tr}([A_a,A_b]^2)$ and of degree three $S_3(A) = \frac{1}{3} \sum_{a,b,c} f_{abc}\mbox{Tr}(A_a[A_b,A_c])$. 

Now suppose we have an estimate $|S_3(A)|^4 \leq \zeta |S_4(A)|^3$ for some $\zeta > 0$ independant of $A$. Then we can use the trick described in \cite{princ_var_rep_th}. For fixed $A$, we consider $S$ on a ray $tA$ and we obtain a real polynomial function of $t$. The minimum of such a function is given by :
\begin{equation}
\label{min_t_S}
\min_t S(tA) = -\frac{|S_3(A)/4|^4}{|S_4(A)/3|^3} \geq -\frac{3^3}{4^4} \zeta \;\;,
\end{equation}
so we get a lower bound for $S$ : $S(A) \geq \frac{3^3}{4^4} \zeta \;\;\; \forall A \;\;$.

To find such an estimate, we use the relation proved in \cite{princ_var_rep_th} for any three anti-hermitian and traceless matrices $A_1$, $A_2$ and $A_3$ :
\begin{equation}
\label{est_al_petr}
(-\mbox{Tr}([A_1,A_2]^2 + [A_2,A_3]^2 + [A_3,A_1]^2))^3 \geq \frac{324}{n^3-n}(\mbox{Tr} A_1[A_2,A_3])^4 \;\;\; .
\end{equation}
The steps of the estimate go as follow :
\begin{equation}
\label{est_S^3}
\begin{array}{rcl}
  \frac{1}{3} |\sum_{abc} f_{abc} \mbox{Tr}(A_a [A_b , A_c])| &\leq& \frac{1}{3}  \sum_{abc} |f_{abc}| |\mbox{Tr}(A_a [A_b , A_c])| \\
  &\leq& \frac{1}{3}  \left ( \frac{n^3-n}{324} \right )^{\frac{1}{4}} \sum_{abc} |f_{abc}| \\
  && (-\mbox{Tr}([A_a,A_b]^2 + [A_b,A_c]^2 + [A_c,A_a]^2))^{\frac{3}{4}} \\
  &\leq& \left ( \frac{n^3-n}{324} \right )^{\frac{1}{4}} \sum_{abc} |f_{abc}| (-\mbox{Tr}[A_a,A_b]^2)^{\frac{3}{4}} \\
  &\leq& \left ( \frac{n^3-n}{324} \right )^{\frac{1}{4}} C \sum_{ab}(-\mbox{Tr}[A_a,A_b]^2)^{\frac{3}{4}} \\
  &\leq& \left ( \frac{n^3-n}{324} \right )^{\frac{1}{4}} C (d(d-1))^{\frac{1}{4}} (-\sum_{ab}\mbox{Tr}[A_a,A_b]^2)^{\frac{3}{4}} \\
  &=& \left ( \frac{n^3-n}{324} \right )^{\frac{1}{4}} C  (d(d-1))^{\frac{1}{4}} (4S_4(A))^{\frac{3}{4}}\;\; .
\end{array} 
\end{equation}

We use (\ref{est_al_petr}) to go from the first line to the second, the antisymmetry properties of the structure constants from the second to the third, and then define $C = \max_{\{a,b\}} \sum_c |f_{abc}|$ when passing to the fourth one. Finally, we use the relation $\sum_{i=1}^m(\alpha_i)^{\frac{3}{4}} \leq (m)^{\frac{1}{4}} (\sum_{i=1}^m \alpha_i)^{\frac{3}{4}}$, valid for any $m$ positive real numbers $\alpha_i$. Therefore we have our estimate and $S(A)$ is always bounded from below.

Let us remark that boundedness of the action from below does not imply the existence of a global minimum. There are actually numerous "valleys" going to infinity in which the value of the action stays bounded. Consider again $S(tA)$ as a real function of $t$ for fixed A on the unit sphere. The minimum of $S$ on the ray $tA$ is given by $S_{min}(A) = -\frac{3^3(S_3(A))^4}{4^4(S_4(A))^3}$. This expression is discontinuous for $S_4(A) = 0$, i.e. when A forms a set of commuting matrices. The minimum in $t$ is achieved at $t_{min}(A) = -\frac{3S_3(A)}{4S_4(A)}$, which happens to diverge on sets of commuting matrices. Diverging $t_{min}$ is the sign of a valley of ray minima going to infinity.

To see a simple example of this phenomenon, consider an irreducible representation $\{B_a\}$ of $\textfrak{g}$, and some set of multiples of the identity $\{C_a\}$. Then $S(B + \gamma C)$ is constant for any real number $\gamma$, and in particular for $\gamma \rightarrow \infty$. Adding a multiple of the identity corresponds to translating the D-brane, and this does not modify the action. These valleys linked with translations are degenerate, so the asymptotic value reached at infinity is also reached at finite distance, but one cannot exclude that there exists a more complicated path approaching a value lower than any value reached at finite distance.

\section{Stability of type II solutions}

\label{sa_rep}

If $A$ is an irreducible representation of $\textfrak{g}$ in $\mbox{Mat}(n)$, some simple computations using the properties of the Casimir operators of the adjoint representation and the $A$ representation give :
\begin{equation}
     \label{S_A_rep_irr}
     S(A) = -\frac{1}{24 h^\vee} n (\lambda, \lambda + 2\rho) \;\;,
\end{equation}
where $\lambda$ is the highest weight of the representation $A$, and $2\rho = \sum_{\alpha>0}  \alpha$ is the sum of the positive roots of $\textfrak{g}$. 

For a simple compact algebra, any representation is a direct sum of irreducible subrepresentations, so these subrepresentations commute with each other. The value of $S(A)$ is then the sum :
\begin{equation}
     \label{S_A_rep_qcq}
     S(A) = -\frac{1}{24 h^\vee} \sum_i n_i (\lambda_i, \lambda_i + 2 \rho) \;\;,
\end{equation}
with $i$ an index running on irreducible subrepresentations, $n_i$ their respective dimensions and $\lambda_i$ their highest weights.

\vspace{.5cm}

Now we turn to the study of stability of such solutions, for which we need the second variation of $S(A)$. The latter can be expressed by an operator $\square$ \nolinebreak:
\begin{equation}
\label{def_square}
\delta^2 S = \frac{1}{4 (h^\vee)^2} \mbox{Tr} \left ( \sum_b\delta A_b \square_b(\delta A) \right )
\end{equation}
\begin{equation}
  \label{d2S_op_gen}
  \Square_{b} (\delta A) = - \left ( [\sigma_a, \sigma_b] - f_{abc} \sigma_c -  \frac{1}{2}\,\delta_{ab} \sigma_c \sigma_c + \frac{1}{2}\,\sigma_b \sigma_a \right ) \delta A_a
  \;\; ,
\end{equation}
where $\sigma_a (B) = [A_a, B]$. When $A$ is a representation, $\sigma : e_a \mapsto \sigma_a$ is also a representation : $[\sigma_a, \sigma_b] = f_{abc} \sigma_c$, so the first two terms cancel. Moreover, as $\sigma_c \sigma_c$ is the Casimir operator $\Gamma_\sigma$ of $\sigma$, the second variation on representations can be written :
$$
  \Square_{b} (\delta A) = \frac{1}{2}(\delta_{ab} \Gamma_\sigma - \sigma_b \sigma_a) (\delta A_a) \;\;.
$$

Notice that a configuration $\{A_a\}$ is a local minimum if all the eigenvalues of $\Square_{b}$ are \emph{negative}, because the bilinear form defined by the trace is negative definite. We consider a matrix-valued eigenvector $X_a$, such that :
\begin{equation}
  \label{eq_val_p}
  \Gamma_\sigma X_b - \sigma_b \sigma_a X_a = 2 \nu X_b \;\;.
\end{equation}
By letting $\sigma_b$ act on this equation and summing on $b$, we get either $\nu = 0$ or $\sigma_b X_b = 0$. The latter is equivalent to $\Gamma_\sigma X_b = 2 \nu X_b$. Projecting both side of this equation on $X_b$ with the trace form, we can check that $\nu \leq 0$ for any representation. We now have to check that the third derivative vanishes in directions such that $\nu$ is zero. If it is the case, we have a local minimum, because the fourth derivative is always positive. Putting $\nu = 0$ in (\ref{eq_val_p}) we get :
\begin{equation}
  \label{val_p_l_0}
  \Gamma_\sigma X_b = \sigma_b \sigma_a X_a \;\;.
\end{equation}

$\Gamma_\sigma$ has a non-zero kernel. For a representation that does not contain any pair of isomorphic subrepresentations, this kernel contains only diagonal matrices, of the form :
\begin{equation}
\label{K_rep_comp_red}
\mbox{Ker}(\Gamma_\sigma) = \left \{ \left (
  \begin{array}{cccc}
    \zeta_1  \mathbbm{1} & 0 & \cdots & 0 \\
    0 & \zeta_2  \mathbbm{1} & \cdots & 0 \\
    \vdots & \vdots & \ddots & \vdots \\
    0 & 0 & \cdots & \zeta_m \mathbbm{1}
  \end{array}
\right ) \;\;:\;\; \zeta_1,\; \zeta_2,\cdots,\;\zeta_m \in \mathbbm{R} \right \} \;\;,
\end{equation}
where the blocks correspond to submodules of dimensions $d_1$, $d_2$,..., $d_m$. Therefore $\Gamma_\sigma$ has an inverse $\Gamma_\sigma^{-1}$ on the space of matrices which are traceless on each block. Let $X_b = Y_b + K_b$, with $K_b \in \mbox{Ker}(\Gamma_\sigma)$ and $Y_a$ traceless on each block. Defining $H = \Gamma^{-1}_\sigma  \sigma_a Y_a$, we get from (\ref{val_p_l_0}) : $X_b = [A_b, H] + K_b$.

Variations of the type $X_b = [A_b, H]$ correspond to transformations $A_a \mapsto U^{-1}A_a U$ with $U = exp(H)$, and $S(A)$ is constant in these directions.

Variations of the type $X_b = K_b$ commute with each other and with the elements of the representation. As explained above, they correspond to shifts of the brane on the Lie group. They leave $S$ constant because of the translation symmetry of the action.

The situation is different when two or more subrepresentations are isomorphic. In this case, there are elements in $\mbox{Ker}(\Gamma_\sigma)$ which are not diagonal, and they do not satisfy simple commutation relations anymore. The third derivative $\delta^3S$ will generally not vanish and such representations are \emph{not} local minima of $S$. Physically, two isomorphic subrepresentations describe two stacked D-branes of the same size, centered at the origin. This configuration is unstable, because such a stack will quickly form a larger single D-brane. This interpretation fits well with the well-known phenomenon of tachyon condensation in SU(2) that turns several D0-branes into a D2-brane \cite{fil12}, and with \cite{fil11}, where it was shown that a D0-brane approaching a D2-brane too close gets absorbed and a larger D2-brane results.

\section{Type III solutions}

\label{sub_rep}

We study now the type III solutions to the equations of motion, i.e. the ones consisting of representations of a subalgebra $\textfrak{h} \in \textfrak{g}$. We choose an orthonormal basis of $\textfrak{g}$ compatible with the decomposition $\textfrak{g} = \textfrak{h} \oplus \textfrak{h}^\bot$. The indices $i,j,k,...$ will denote directions tangent to $\textfrak{h}$, while indices $r,s,t,...$ are assigned to directions perpendicular to $\textfrak{h}$. Indices $a,b,c,...$ still run over the whole basis of $\textfrak{g}$.

Let us compute first the value of the action on such solutions. We choose matrices $\{A_i\}$ forming a representation of $\textfrak{h}$, and set $\{A_s\}$ to zero. In $\textfrak{g}$, we have $[e_i,e_j] = f_{ijk}e_k$. Now if we choose a basis $\{e'_i\}$ of $\textfrak{h}$ orthonormal with respect to the Killing form of $\textfrak{h}$, the structure constants relative to this new basis will differ by a common constant $\mu$, which will depend on the embedding of $\textfrak{h}$ in $\textfrak{g}$ : $[e'_i,e'_j] = f'_{ijk}e'_k$ with $f'_{ijk} = \mu f_{ijk}$. $\mu^2$ is sometimes called the \emph{embedding index} of the subalgebra (see \cite{conf_f_th}, §13.7). 

We introduce matrices $B_i = \mu A_i$, which satisfy $[B_i, B_j] = f'_{ijk} B_k$ and the action can now be written :
\begin{equation}
  \label{S_srep}
  \begin{array}{rl}
    S_{\textfrak{g}}(A) &= \frac{1}{4(h^\vee_g)^2} \left (-\frac{1}{4} \sum_{i,j}        \mbox{Tr}([A_i,A_j]^2) + \frac{1}{3} \sum_{i,j,k} f_{ijk} \mbox{Tr}(A_i[A_j,A_k]) \right ) \\
    &= \frac{1}{4(h^\vee_g)^2} \frac{1}{\mu^4} \left (-\frac{1}{4} \sum_{i,j} \mbox{Tr}([B_i,B_j]^2) + \frac{1}{3} \sum_{i,j,k} f'_{ijk} \mbox{Tr}(B_i[B_j,B_k]) \right ) \\
    &= \frac{(h^\vee_h)^2}{\mu^4(h^\vee_g)^2}S_{\textfrak{h}}(B) \;\;.
  \end{array}
\end{equation}
We added subscripts to the actions and to the Coxeter numbers to indicate to which Lie algebra they refer. Up to the factor $\frac{(h^\vee_h)^2}{\mu^4(h^\vee_g)^2}$, the value of the $\textfrak{g}$-action on representations of a subalgebra $\textfrak{h}$ is equal to the $\textfrak{h}$-action. Using (\ref{S_A_rep_qcq}), this allows to compute $S$ on any type III solution, provided we can determine the factor $\mu$.

As we will see later, the lowest energy configurations appear for $\textfrak{h} =$ su(2), so we will concentrate our attention on three-dimensional subalgebras (TDS) of $\textfrak{g}$. In any TDS of $\textfrak{g}$ we can choose generators $\hat{e}_+,\hat{e}_-,\hat{e}_3$ such that $[\hat{e}_3,\hat{e}_+]=2\hat{e}_+$, $[\hat{e}_3,\hat{e}_-]=-2\hat{e}_-$ and $[\hat{e}_+,\hat{e}_-]=\hat{e}_3$. Let $\textfrak{g}_0$ a Cartan subalgebra of $\textfrak{g}$ containing $\hat{e}_3$. The element $\phi \in \textfrak{g}_0^*$ dual to $\hat{e}_3$ by the Killing form is called the \emph{defining vector} of the TDS, and any two TDS with Weyl conjugate defining vectors are equivalent \cite{dyn_sub}. We can compute the norm of $\phi$ and get \nolinebreak:
$$
\left\|\phi\right\| = \left\|\hat{e}_3\right\| = \sqrt{2} \left\|e'_3\right\| = \sqrt{2} \mu \left\|e_3\right\| = \sqrt{2} \mu \;\;,
$$
which allows to determine the parameter $\mu$.

$S$ is negative on representations, so we are looking for the minimal value of $\mu$ that will minimize (\ref{S_srep}). There is a simple argument showing that $\mu \geq 1$. The adjoint action of any TDS splits $\textfrak{g}$ into irreducible su(2) modules. At least one copy of the adjoint (spin 1) module is present, as it consists of the subalgebra itself. Now the $\hat{e}_3$-eigenvalue of a generator $e_\alpha$ of $\textfrak{g}$ associated with the root $\alpha$ is given by $(\phi,\alpha)$, and this must be equal to 2 for at least one of the $e_\alpha$. Combining this with the condition that $\left\|\phi\right\|$ should be minimal, we conclude that $\phi$ must be equal to a long root, in which case $\mu = 1$. This defines the three dimensional minimal embedding index subalgebras (MIS). They are all equivalent, because all the long roots are Weyl equivalent. It will be useful to notice that under the action of MIS, $\textfrak{g}$ decomposes into the following sum of su(2) modules : $(1) \oplus k(\frac{1}{2}) \oplus l(0)$. We denoted su(2) modules by their spin between parenthesis and included $k,l$ the multiplicities of the fundamental and trivial modules. The crucial fact is that the adjoint appears only \emph{once}, and that no higher spin module is present. 

As an example of a MIS, consider su($N$), and the subalgebra generated by matrices having the following block form :
$$
\left (
\begin{array}{cc}
\tau_i & 0 \\
0&0
\end{array}
\right ) \in \mbox{su}(N) \;\;,
$$
where $\tau_i$ are generators of the fundamental 2-dimensional su(2) representation. This a MIS, and under its adjoint action, su($N$) decomposes into the following sum of su(2) modules \nolinebreak: $(1) \oplus 2(N-2)(\frac{1}{2}) \oplus (N-2)^2 (0)$.

\section{Stability of MIS representations}

\label{stab_srep}

We keep the same convention on indices. Using the fact that $f_{ijs} = f_{sij} = f_{jsi} = 0$, the second derivative operator can be written on representations of subalgebras as (see (\ref{d2S_op_gen})) :
$$
\Square_{i} (\delta A) = \frac{1}{2}(\delta_{ij} \sigma_k \sigma_k - \sigma_i \sigma_j) \: \delta A_j
$$
\begin{equation}
\label{secderorth}
\Square_{r} (\delta A) = \left (f_{tri} \sigma_i +  \frac{1}{2}\,\delta_{rt} \sigma_i \sigma_i \right ) \delta A_t \;\;,
\end{equation}
where we defined as before $\sigma_i(B) = [A_i,B]$. Note that the second derivative bilinear form does not mix $\delta A_i$'s with $\delta A_r$'s. For variations of the generators of the representation ($\delta A_i$), it is the same as for a full representation of the algebra, so the same conclusions apply : only subalgebra representations having no isomorphic subrepresentations can be local minima. We still have to study what happens with the variations of the matrices set to 0 ($\delta A_r$). We will now restrict ourselves to the configuration giving the lowest value for the action : an irreducible representation of the su(2) MIS.

So suppose $\delta A_i = 0$ and look only at the second set of equations of (\ref{secderorth}). We consider the complexified Lie algebra $\mathbbm{C}\textfrak{g}$, and choose a basis compatible with the decomposition of $\mathbbm{C}\textfrak{g}$ into sl(2) modules under the action of the MIS. We can take orthonormal generators in the sl(2) MIS, but the whole basis will not be orthonormal in general. As $\Square$ respects the module structure, we consider only variations associated to a single sl(2) module. We denote it by $V_S$, with $S$ its spin. Because the remark at the end of the previous section, $S$ can be equal either to 0 or to $\frac{1}{2}$.

We look at $\delta A$ as an element of $V_S \otimes Mat(n)$, and we rewrite $\Square$ as :
$$
\Square = - \sum_i \tau(e_i) \otimes \sigma_i - \frac{1}{2} 1 \otimes \Gamma_\sigma \;\;,
$$
where $\tau$ is the representation of the MIS in the module $V_S$. Recalling that the Casimir operator of the tensor representation $\tau \otimes \sigma$ is given by :
$$
\Gamma_{\tau \otimes \sigma} = \sum_i (\tau(e_i) \otimes 1 + 1 \otimes \sigma_i)^2 \;\;,
$$
this equation can be expressed in term of Casimir operators only :
$$
\Square = - \frac{1}{2} (\Gamma_{\tau \otimes \sigma} - \Gamma_\tau \otimes 1) \;\;.
$$

Under the action $\sigma$ of the irreducible sl(2) representation $A$, Mat($n$) decomposes into a direct sum of modules of integer spin $W_s$, $s = 0,1,\ldots,n-1$. $\Gamma_\tau$ acts on a module of spin $S = 0$ or $\frac{1}{2}$. $\Gamma_{\tau \otimes \sigma}$ acts on modules of spin $s' = s$ if $S = 0$ and of spin $s' = s\pm\frac{1}{2}$ if $S = \frac{1}{2}$. So from the equation above, we conclude that $\Square$ has strictly negative eigenvalues, except in two cases, where null directions exist : $(s = 0)$ and $(s=1,S=\frac{1}{2})$. We now relate these null directions to symmetries of S.

For $s=0$, the variations $\delta A_r$ have zero $\Gamma_\sigma$ eigenvalue, and are therefore multiples of the identity, because $A$ is an irreducible representation. We already saw that such variations leave the action constant, because of its translation invariance.

For the case $(s=1,S=\frac{1}{2})$, the tensor product of modules $V_{\frac{1}{2}} \otimes W_1$ splits under the action of $\tau \otimes \sigma$ into one spin $\frac{3}{2}$ module $T_{\frac{3}{2}}$ and one spin $\frac{1}{2}$ module $T_{\frac{1}{2}}$. $\Square$ has a negative eigenvalue on $T_{\frac{3}{2}}$, but zero eigenvalue on $T_{\frac{1}{2}}$. Using the basis $\{A_i\}$ in $W_1$ and choosing a basis $\{e_\uparrow,e_\downarrow\}$ in $V_{\frac{1}{2}}$, we can compute $\Gamma_{\tau \otimes \sigma}$ explicitly and find the two generators of $T_{\frac{1}{2}}$, on which $\Square$ vanishes. They are given in the following table by their components on the product basis :
$$
\begin{array}{cccccccc}
&\stackrel{1}{\scriptstyle \uparrow} &\stackrel{2}{\scriptstyle \uparrow}&\stackrel{3}{\scriptstyle \uparrow}&\stackrel{1}{\scriptstyle \downarrow}&\stackrel{2}{\scriptstyle \downarrow}&\stackrel{3}{\scriptstyle \downarrow}& \\
(&-i&1&0&0&0&i&) \\
(&0&0&i&i&1&0&) \\
\end{array} \;\;.
$$

In the remaining of this section, we want to show that these null directions are linked with the invariance of $S$ under conjugation action of $G$ on $\textfrak{g}$. Actually, we saw that the MIS is unique only up to equivalence, i.e. up to conjugation with elements of the group $G$. Let us check that these transformations correspond to the null directions found above. For an infinitesimal transformation of this type, the variation of $e_a$ is given by $\delta e_a = [h, e_a]$ where $h$ is the infinitesimal antihermitian generator of the transformation. The variation of the map $A$ is then $\delta A_r = (\delta e_i, e_r) A_i $. We introduce $h_\uparrow = (h, e_\uparrow)$ and $h_\downarrow = (h, e_\downarrow)$, the components of $h$ on $V_{\frac{1}{2}}$. Then \nolinebreak:
$$
(\delta e_i, e_r) = ([h, e_i], e_r) = (h, [e_i, e_r]) = \overline{f_{ir}^{\;\;s}}h_s \;.
$$
The structure constants $f_{ir}^{\;\;s}$ are determined by the representation $\tau$ of the MIS on $V_{\frac{1}{2}}$ :
\begin{equation}
\label{actionMIS}
\tau(e_1) = \frac{1}{\sqrt{2}} \left ( 
\begin{array}{cc} 
0 & i \\ i & 0
\end{array}
\right ) ,\;
\tau(e_2) = \frac{1}{\sqrt{2}} \left ( 
\begin{array}{cc} 
0 & -1 \\ 1 & 0
\end{array}
\right ) ,\;
\tau(e_3) = \frac{1}{\sqrt{2}} \left ( 
\begin{array}{cc} 
i & 0 \\ 0 & -i
\end{array}
\right ).
\end{equation}
Setting $(h_\uparrow, h_\downarrow)$ to ($\sqrt{2}$,0) and (0,$-\sqrt{2}$), we find the two eigenvectors in the table above. 

So the null directions found in the $(s=1,S=\frac{1}{2})$ case do correspond to the invariance of S under the conjugation action of $G$ on $\textfrak{g}$. These transformations can be seen as changes of basis in $\textfrak{g}$, and physically, they are rotations of the fuzzy sphere D2-brane in the d-dimensional group manifold.

Hence we have proved that irreducible representations of MIS are local minima of $S$. 
\section{Some examples in su(3) and beyond}

\label{exam}

Now we would like to use the work above to do some phenomenology and compare the energy of the various brane configurations in the next simplest case beyond su(2) : su(3).

We first specialize the formulas (\ref{S_A_rep_irr}) and (\ref{S_srep}) for su(3). For $A$ a full su(3) representation, characterized by it's Dynkin indices $(p,q)$, we have :
\begin{equation}
\label{S_A_rep_su3}
S(A_{(p,q)}) = -\frac{1}{216}(p+1)(q+1)(p+q+2)(p^2 + 3p + pq + 3q + q^2)\;\;.
\end{equation}
For $\{A_i\}$ forming a representation of the MIS and $A_r = 0$ we get :
\begin{equation}
\label{S_A_rep_su2_in_su3}
S(A_{(p)}) = -\frac{1}{216}p(p+1)(p+2)\;\;,
\end{equation}
with $p$ the Dynkin index of the su(2) representation (twice the spin).

We can make a little table for the value of $S$ on these solutions for various $n$ :
\begin{center}
\begin{tabular}{|c|c|c|c|c|}
 \hline $n$ & su(3) Representation & $S$ su(3) & MIS Representation & $S$ MIS \\
 \hline  3  & $(1,0)$   & -4/36   & (2) & -4/36 \\
 4 & $(1,0) \oplus (0,0)$ & -4/36 & (3) & -10/36 \\
 5 & $(1,0) \oplus 2(0,0)$ & -4/36 & (4) & -20/36 \\
 6 & $(2,0)$ & -20/36 & (5) & -35/36 \\
 7 & $(2,0) \oplus (0,0)$ & -20/36 & (6) & -56/36 \\
 8 & $(1,1)$ & -24/36 & (7) & -84/36 \\
 \hline
\end{tabular}
\end{center}
The columns give respectively the size of the Chan-Patton gauge group ($n$), the representation of su(3), the value of S on the latter, the su(2) representation, and again S computed on the latter. We observe that the MIS solutions seem to be systematically lower in energy, except for $n = 3$. This contradicts the conjecture made in \cite{princ_var_rep_th}, and the general belief that branes on Lie group adopt the maximally symmetric geometry of conjugacy classes.

Passing by, we can point out an interesting phenomenon. If we consider only maximally symmetric solutions, it's not always true that irreducible representations have lower energy than reducible ones, contrary to the su(2) case. For instance if we choose $n = 27$, the irreducible representation $(2,2)$ gives $S = -18/3$, whereas $(5,0) \oplus (2,0)$ gives $S = -25/3$. This is due to the fact that, in su($N$), representations with weight $\lambda = (m, 0, \ldots, 0)$ are the ones which maximize the ratio $\left\|\lambda\right\|/n$. They give lower value for the action compared to other representations with the same $n$. So maximally symmetric D-brane are sometimes more stable when split than when they are completely condensed. Anyway in this case the value of the MIS solution is $-91 = -273/3$ (!).

\vspace{.5cm}

We give some arguments to show that fuzzy sphere D2-branes have lower energy that maximally symmetric ones in most cases. We will consider the action $S$ for $G=$ SU($N$), $N \geq 3$, and compare its value on an irreducible MIS solution and on a representation of the full algebra (type II). We will choose the weight for the latter of the form $\lambda = (m,0,\ldots, 0)$, as these give the lowest action value among full representations, as we hinted above.

We first use Weyl dimension formula to get an expression for the dimension $n$ of the representation as a function of $m$ :
\begin{equation}
\label{n(m)}
n = \prod_{\alpha > 0} \frac{(\lambda + \rho,\alpha)}{(\rho,\alpha)} = \prod_{i = 1}^{N-1} \frac{m+i}{i} \;\;.
\end{equation}
Using (\ref{S_A_rep_irr}), we get for the full representation :
$$
S_{\mbox{\tiny full}} = -\frac{1}{24N}n \left (\frac{N-1}{N}m^2+(N-1)m \right ) \;\; ,
$$
and using (\ref{S_srep}), for the MIS representation :
\begin{equation}
\label{S_Mis}
S_{\mbox{\tiny MIS}} = -\frac{n^3-n}{24N^2} \;\;.
\end{equation}
We now have to compare these results :
$$
\begin{array}{rclc}
\frac{n^3-n}{N} & \stackrel{?}{\geq} & n\left (\frac{N-1}{N}m^2+(N-1)m \right ) & \Leftrightarrow \\
\frac{n^2}{N(N-1)} & \stackrel{?}{\geq} & \frac{m^2}{N}+m+\frac{1}{N(N-1)}\;\; . &
\end{array}
$$
With (\ref{n(m)}), one can check that provided $m > 1$, the left-hand side above is an increasing function of N, while the right hand side is obviously a decreasing one. So it's sufficient to check the inequality when $N = 3$ for all $m > 1$, which is trivial when writing $n$ explicitly. When $m = 1$, we have equality for all $N$. Therefore $S_{\mbox{\tiny full}} \geq S_{\mbox{\tiny MIS}}$, with equality only in the case $n = N$. So the fundamental representation of su($N$) has the same value as the MIS solution, but else the latter is always lower in energy.

As pointed out above, the special type of representations of su($N$) we considered usually gives the lowest value of the action, even if a complete proof of this fact is lacking. So the comparison above is at least a strong argument suggesting that MIS solutions are lower in energy than type II ones. This also justifies a posteriori our choice to focus our attention on su(2) subalgebras, rather than on larger ones.

Finally, still considering su($N$), we can also compare the MIS solutions with the lower bound found in section \ref{ex_min}. We replace $\zeta$ in (\ref{min_t_S}) by its explicit value found in (\ref{est_S^3}) and, introducing again the factor $\frac{1}{4(h^\vee)^2}$, we get :
$$
S \geq -\frac{n^3-n}{192N^2} C^4 d(d-1) \;\;.
$$
Comparing with (\ref{S_Mis}), the value of the action for a MIS has the same behaviour in $n$ as the lower bound, which is not the case for representations of larger subalgebra or of the full algebra. 

The arguments above lead us to think that the global minimum is reached on irreducible representations of MIS, and therefore that the brane ground state is a fuzzy sphere D2-brane in any compact simple Lie group.

\section{Conclusion}

In this note, we proved the boundedness of the action (\ref{def_S}) for any simple compact Lie group $G$. 

We saw that a given configuration of the gauge field $A$ forming a representation of the Lie algebra $\textfrak{g}$ of $G$ is a local minimum if and only if it does not contain any pair of isomorphic subrepresentations. Physically, these configurations correspond to sets of maximally symmetric D-branes centered at the origin, and the unstable configurations arise when two or more D-branes are stacked.

We gave a formula to compute the value of the action when the gauge field is a representation of a proper subalgebra of $\textfrak{g}$, and hence breaks the Kac-Moody symmetry of the boundary currents. We proved that irreducible representations of a special type of three dimensional subalgebras, namely the ones with minimal embedding index, are local minima.

For $G =$ SU(3), we also computed explicitly the value of the action on su(3) representations of various dimensions, as well as on MIS representations. We showed that the latter take lower values of $S$ than su(3) representations. In the D-brane picture, this means that maximally symmetric D-branes are not stable in SU(3). Fuzzy sphere D2-branes are definitely much lower in energy, but the question about whether they are really the ground state remains open.

We gave some arguments to show that the result still holds for SU($N$). We saw that the behaviour in $n$ of the MIS solutions is the same as the lower bound of the action found before, which is one more incentive to think they can correspond to the brane ground state.

\vspace{.3cm}

The geometrical differences between inequivalent TDS representation solutions are still unclear, and in particular we do not know if there is some geometrical feature that would distinguish the most stable MIS D2-brane from other TDS D2-branes. 

Fuzzy sphere D2-branes seem to appear in every compact Lie group when they are small compared to the radius of curvature. Interestingly, the dimension of the group manifold has no influence on the dimension of the brane. We can finally remark that in the limit we considered, the D-brane "sees" only a small piece of a manifold with positive curvature. So one can wonder whether this brane configuration really depends on the structure of Lie group, and whether it could not be a stable form for D-branes in any positively curved part of a generic manifold.

\vspace{0.5cm}

{\bf Acknowledgment} : I would like to thank very much Anton Alekseev, for initiating this work and for his constant support. Thanks go also to Rudolf Rohr for useful discussions about three dimensional subalgebras. This research was supported in part by the Swiss National Science Foundation.

\end{document}